\documentclass[a4,prd, floatfix, nofootinbib,
preprintnumbers
,twocolumn
]{revtex4}
\usepackage[utf8x]{inputenc}
\usepackage[T1]{fontenc}
\usepackage{lmodern}
\usepackage{datetime}
\usepackage{amsmath}
\usepackage{amssymb}
\usepackage{bm}
\usepackage{epsfig}
\usepackage{graphicx}
\usepackage{color}
\usepackage{hyperref}
\usepackage{slashed}

\newcommand{\be}{\begin{equation}}
\newcommand{\ee}{\end{equation}}
\newcommand{\bdm}{\begin{displaymath}}
\newcommand{\edm}{\end{displaymath}}
\newcommand{\bea}{\begin{eqnarray}}
\newcommand{\eea}{\end{eqnarray}}

\newcommand{\cor}[1]{{\color{black}#1}}

\newcommand{\ipnp}{Institute of Particle and Nuclear Physics,
  Faculty of Mathematics and Physics,
  Charles University in Prague, V Hole\v{s}ovi\v{c}k\'ach 2,
  180 00 Praha 8, Czech Republic}

\begin{document}
\title{Resurgent structure of the 't~Hooft-Polyakov monopole}
\preprint{}
\pacs{\bf PACS should be added at some point}
\author{Michal Malinsk\'{y}}
\email{michal.malinsky@matfyz.cuni.cz}
\affiliation{\ipnp}
\begin{abstract}
In this letter we present a comprehensive analysis of the differential equations governing the spatial profile of the 't~Hooft-Polyakov monopole from the viewpoint of resurgence theory. We note that the universality of the gauge-component asymptotics, together with the relative simplicity of its Borel transform and the associated Volterra equations' kernels, gives rise to a perturbative expansion  featuring a good control over the proliferation of the Borel-plane singularities to all orders, along with full information about the relevant logarithmic discontinuities. Moreover, its partial resummation reveals remarkably simple universal analytic non-perturbative background profiles, around which one can develop a uniformly convergent global perturbative expansion of the exact solutions for any $\lambda/e^2>0$. This also provides an analytic grip on the numerical parameters governing the expansions of both the gauge and scalar profile functions at the origin and at infinity. 
%\\{\bf As of {\small \today\; \currenttime}}   
\end{abstract}
\maketitle
\section{Introduction\label{sec:intro}}
It is well known that non-linear ordinary differential equations (ODEs) with domains stretching all the way to $+\infty$ typically do not admit convergent large-$x$ transseries expansions and, with the notable exception of the BPS case~\cite{Prasad:1975kr,Bogomolny:1975de}, the ODEs governing the spatial profile of the 't~Hooft-Polyakov monopole~\cite{tHooft:1974kcl,Polyakov:1974ek},
\begin{eqnarray}\label{eq:gaugeprofile}
y''&=&y \,z^2 +y(y^2-1)/x^2\,,\\
\label{eq:scalarprofile}
z''+2z'/x &=&2 z \,y^2/x^2+\beta z(z^2-1)\,,
\end{eqnarray}
(with  $\beta\equiv \lambda/e^2$, $x$ corresponding to $r$, $y$ to $K$ and $z x$ to $H$ of ref.~\cite{Prasad:1975kr}) are no different in this respect.
Hence, the usual approach to their solution exploits various types of numerical accounts based on controlled small-parameter or local expansions, or even vanilla ODE integrators like Runge-Kutta (cf. \cite{Julia:1975,Bais:1976,Goddard:1977da,Gardner:1983,Breitenlohner:1992,Forgacs:2005vx} and references therein).  

What we claim in this letter is that even the basic methods of resurgence theory (cf.~\cite{Ecalle:1981,Sternin:1996,Costin:2008,Aniceto:2011nu,Dunne:2013ada,marinoResurgenceNotes,Aniceto:2013fka,Dorigoni:2014hea,Sauzin:2016,Aniceto:2018bis,Dunne:2025mye}) make it possible to do much better in this respect. The method is the full exploitation of the analytic form of the lowest exponential level gauge-sector (\ref{eq:gaugeprofile}) asymptotics Borel resum, and a careful bookkeeping of the proliferation of singularities throughout higher levels of the relevant Volterra equations.      
The key to this is the fact that, due to the exponentially fast saturation of the {\nolinebreak $z|_{x\to \infty}\to 1$} boundary condition in the $\beta>0$ regime, the asymptotics of solutions of the gauge profile ODE~(\ref{eq:gaugeprofile}) is universal
\begin{equation}\label{eq:yasymptotics}
y(x\to +\infty)\to A_\beta \sqrt{x}K_\nu(x)\qquad \forall \beta>0\,,
\end{equation} 
(up to a $\beta$-dependent real constant $A_\beta$); here $K_\nu$ denotes the modified Bessel $K$-function of imaginary order $\nu=i \sqrt{3}/2$.
At the same time, the $\sqrt{x}K_\nu(x)$ piece of Eq.~(\ref{eq:yasymptotics}) can be cast in terms of a Laplace transform of an associated ${}_2F_1$ hypergeometric function (see e.g.\cite{Dunne:2025mye}),
\begin{equation}\label{eq:KBorelexpression}
\sqrt{x}K_\nu(x)\!=\!\sqrt{\tfrac{\pi}{2}}\,x e^{-x}\int_0^\infty\!\!\!\! e^{-x t}{}_2F_1(\tfrac{1}{2}\!-\!\nu,\tfrac{1}{2}\!+\!\nu;1;-\tfrac{t}{2})\,,
\end{equation}
where the pre-integral $x e^{-x}$ factor is fully compatible with the \cor{inverse-}power-exponential shape of the relevant asymptotic transseries, 
as well as with that of the asymptotically fast decaying $y^3/x^2$ non-linearity of Eq~(\ref{eq:gaugeprofile}). 
Hence, the hypergeometric integrand of Eq.~(\ref{eq:KBorelexpression}) \cor{can be viewed as a seed of} the Borel-plane structure of solutions of Eqs.~(\ref{eq:gaugeprofile})-(\ref{eq:scalarprofile}) regardless of the value of $\beta>0$.

Interestingly, the discrete singularity structure of $\hat{f}_0(t)=A_\beta\sqrt{\tfrac{\pi}{2}}{}_2F_1(\tfrac{1}{2}-\nu,\tfrac{1}{2}+\nu;1;-\tfrac{t}{2})$ of Eq.~(\ref{eq:KBorelexpression}), limited to a single branch cut at $t=-2$, along with the triangular (and, hence, iterative) nature of the Borel-plane Volterra equations equivalent to Eqs.~(\ref{eq:gaugeprofile})-(\ref{eq:scalarprofile}), make it possible to trace the resurgent structure of the $y$ and $z$ functions to any desired level -- higher-order singularities are generated from the lower-order ones at very specific points along the real-$t$ axis, inheriting all their structure from the fundamental $\hat{f}_0$ seed. This implies that it should be possible to write the full forms of $y$ and $z$ in the corresponding perturbative sector in terms of Laplace transforms of the full collection of Borel-plane germs with all singularities (whenever encountered) accounted for by median resummation, and the system should also support a well-behaved Borel-Pad\'e-Laplace approach, (cf.~\cite{Dunne:2025mye}).

Even more profoundly, as we shall see, thus obtained Borel-plane perturbation series can be rearranged (and partially resummed) in such a way to provide a {\em uniformly convergent global expansion scheme developed around analytic non-perturbative background profiles} driven by very simple resummed vector-sector ``seed functions'' $\hat{f}_0^R$ and $\hat{g}_0^R$, that yield $x$-plane backgrounds  automatically obeying all boundary conditions. These, in turn, provide an analytic grip on the a-priori unknown parameters $A_\beta$ (and $B_\beta$) governing the asymptotic $x\to \infty$  (and locally convergent $x\to 0$) expansions of both $y$ and $z$.     

In what follows, we shall first demonstrate these principles on the simplified case of the ``maximally non-BPS'' (MNBPS) monopole corresponding to $\beta\to \infty$, where only the gauge ODE~(\ref{eq:gaugeprofile}) retains a non-trivial form (Sec.~\ref{sec:MNBPS}). In Sec.~\ref{sec:general} we shall comment on how the vector sector feeds into the structure of the scalar part in case of general $\beta>0$ configurations. The partial resummation of the na\"\i ve expansion around $\hat{f}_0$ and the explicit construction of the nonperturbative background/core profile $\hat{f}^R_0$, along with a sample calculation of the lowest-order contribution to the $B_\infty$ parameter, are discussed in Sec.~\ref{sec:resummation}. Most of the technical details concerning namely the finite $\beta>0$ (NBPS) case are deferred to a set of Appendixes and an extended study~\cite{MalinskyFuture}.

%%%%%%%%%%%%%%%%%%%%%%%%%%%%%%%%%%%%%%%
\section{The MNBPS monopole ($\beta\to\infty$)\label{sec:MNBPS}}
The variant of Eq.~(\ref{eq:gaugeprofile}) governing the gauge-field profile of the MNBPS monopole, subject to boundary conditions $y|_{x\to 0}\to 1$ and 
$y|_{x\to \infty} \to  0$, reads
\begin{equation}\label{eq:MNBPSprofile}
y''=y+(y^3-y)/x^2\,.
\end{equation}
Despite its apparent simplicity (note the triviality of the crosstalk with the scalar sector $z(x)=1$ $\forall x>0$), Eq.~(\ref{eq:MNBPSprofile}) is still notoriously hard (second order, non-linear, non-conservative, with boundary conditions as limits in the two singular points of its domain).

The Borel-plane representation of the solution~(\ref{eq:KBorelexpression}) of its linearized form, cf. Appendix~{\ref{app:y0MNBPS}}, suggests a substitution 
$y(x)= x e^{-x}f(x)$, 
for which Eq.~(\ref{eq:MNBPSprofile}) assumes a simple symbolic form
$L_\infty[f](x)=e^{-2x}f(x)^3$. 
Here
\begin{equation}\label{eq:foperator}
L_\infty[f](x)\equiv f''(x)- 2 f'(x)+\frac{2f'(x)}{x}-\frac{2f(x)}{x}+\frac{f(x)}{x^2}   
\end{equation}
is the differential operator of the linear problem, and the RHS is a fast-decaying non-linear forcing term. 
In the Borel plane, Eq.~(\ref{eq:MNBPSprofile}) is thus equivalent to
\begin{equation}\label{eq:Volterra}
t(t+2)\hat{f}(t)+\int_0^t {\rm d}s\, K(t,s)\hat{f}(s)=[\hat{f}*\hat{f}*\hat{f}](t-2)\,,
\end{equation}
where 
$K(t,s)=t-3s-2$ is a regular kernel corresponding to the non-derivative part of the differential operator~(\ref{eq:foperator}), and the star symbols in the square bracket stand for (associative and commutative) Borel convolutions
\begin{displaymath}
[\hat{h}_1*\hat{h}_2](t)\equiv \int_0^t {\rm d}s\, \hat{h}_1(s)\hat{h}_2(t-s)\,.
\end{displaymath}
%======================================
\subsection{Proliferation of singularities of $\hat{f}$}\label{sec:singularities}
The Borel-plane picture of the solution of the linearized version of Eq.~(\ref{eq:MNBPSprofile}) written in terms of $\hat{f}_0(t)$ (see Sec.~\ref{sec:intro}) trivially obeys Eq.~(\ref{eq:Volterra}) with zero RHS and, as such, it represents the initial level of the ladder of approximations to the full $\hat{f}=\hat{f}_0+\hat{f}_1+\ldots$. Higher orders are then obtained by iterations exploiting the triangular structure of Eq.~(\ref{eq:Volterra}), for instance
\begin{displaymath}
t(t+2)\hat{f}_1(t)+\int_0^t {\rm d}s\, K(t,s)\hat{f}_1(s)=[\hat{f}_0*\hat{f}_0*\hat{f}_0](t-2)\,,
\end{displaymath}
and so on.
Throughout these, proliferation of the primary $t=-2$ singularity of $\hat{f}_0$ is governed by two simple rules: ii) Convolutions generate new singularities at points corresponding to the shifts by position-vectors of singularities of individual components. ii) The $(t-2)$ argument on the RHS of Eq.~(\ref{eq:Volterra}) then shifts all these by two units to the right. 
Hence, the singularities of $\hat{f}$ will eventually populate the real axis of the Borel plane at discrete $t_{\rm sing.}=2m$ ($m\in \mathbb{Z}$) points.  

%===========================
\section{The general NBPS case with $\beta\in \mathbb{R}_+$\label{sec:general}}
Remarkably enough, the same singularity pattern emerges even for finite $\beta>0$. Indeed, with the substitutions $y(x)=x e^{-x}f(x)$ and $z(x)=1-g(x)$, both $f$ and $g$ can be expanded into asymptotic series with $e^{-2 n x}$ towers, since Eqs.~(\ref{eq:gaugeprofile})-(\ref{eq:scalarprofile}) in these coordinates read
\begin{eqnarray}\label{eq:ODEGeneralfg}
L_f[f]&=&e^{-2x}f^3+ g(g-2)f\,,\\
L_g[g]&=&2e^{-2x}(g-1)f^2+\beta g(g-1)(g-2)\,,\nonumber
\end{eqnarray}
with $L_f[f]=L_\infty[f]$ of Eq.~(\ref{eq:foperator}) and $L_g[g](x)=g''(x)+2\tfrac{g'(x)}{x}$. Note also that the MNBPS limit of Sec.~\ref{sec:MNBPS} is  achieved for $g(x)=0$ which corresponds to $z(x)=1$.
With this at hand, the relevant Volterra equations can be written readily:
\begin{eqnarray}\label{eq:VolterraGeneral}
\hat{L}_f[\hat{f}](t)\!&=&\![\hat{f}*\hat{f}*\hat{f}](t-2)-2[\hat{g}*\hat{f}](t)+[\hat{g}*\hat{g}*\hat{f}](t)\,,\nonumber\\
\hat{L}_g[\hat{g}](t)\!&=&\!-2[\hat{f}*\hat{f}](t-2)+2[\hat{g}*\hat{f}*\hat{f}](t-2)\nonumber\\
&  &+\beta[2\hat{g}-3\hat{g}*\hat{g}+\hat{g}*\hat{g}*\hat{g}](t) \,,\label{eq:fdrivingg}
\end{eqnarray}
with $\hat{L}_f[\hat{f}](t)$ corresponding to the LHS of Eq~(\ref{eq:Volterra}) and $\hat{L}_g[\hat{g}](t)=t^2\hat{g}(t)+\int_0^t {\rm d}s\, \tilde{K}(t,s)\hat{g}(s)$, with $\tilde{K}(t,s)=-2s$.
The system~(\ref{eq:VolterraGeneral}) is very interesting for several reasons: i)~Due to $g(x)|_{x\to\infty}\to 0$, cf. Eq.~(\ref{eq:ODEGeneralfg}), the Borel-plane asymptotic scalar profile is trivial ($\hat{g}_0(t)=0$), and the asymptotic gauge profile $\hat{f}_0(t)$ is still proportional to ${}_2F_1(\tfrac{1}{2}-\nu,\tfrac{1}{2}+\nu;1;-\tfrac{t}{2})$ like in the MNBPS case of Sec.~\ref{sec:MNBPS} (albeit with different normalization factors $A_\beta\neq A_\infty$, cf. Sec.~\ref{sec:intro}); ii)~The proliferation of singularities due to~(\ref{eq:VolterraGeneral}) follows the same pattern as before -- all of them  are seeded by the primary branch point of $\hat{f}_0(t)$ at $t=-2$, the RHS convolutions combine these and, due to the $t\to t-2$ shifts, spread them equidistantly along the real axis (to $t_{\rm sing.}=2m$; $m\in \mathbb{Z}$); iii)~{\em The scalar sector is completely ``enslaved'' by the gauge one} -- nonzero contributions to $\hat{g}_k$'s are generated only from $\hat{f}_k$'s through the first convolution on the RHS of Eq.~(\ref{eq:fdrivingg}).
Due to this, both $f(x)$ and $g(x)$ should have convergent expansions around $x=0$  (and asymptotic ones in the $x\to \infty$ domain), that can be determined by comparison of coefficients in Eqs.~(\ref{eq:ODEGeneralfg}), see also Appendixes~\ref{app:MNBPStransseriesInfinity} and~\ref{app:MNBPStransseries}. %Moreover, one may expect eventual cancellation of Frobenius logs at $x=0$ out of the MNBPS $(\beta\to\infty)$ limit (cf.~\cite{Dunne:2026hfx}). 
This, however, is beyond the scope of this letter and will be fully elaborated on in the extended study~\cite{MalinskyFuture}.  
%%%%%%%%%%%%%%%%%%%%%%%%%%%%%%%%%%%
\section{Partial resummation of $\hat{f}$\label{sec:resummation}}
Concerning the promised lowest-order calculation of the $B_\infty$ parameter of the local power-log expansion of the MNBPS gauge profile around $x=0$, cf. Sect.~\ref{sec:intro}, unfortunately, the background profiles $\hat{f}_0$ and $
\hat{g}_0$ of Sections~\ref{sec:MNBPS} and~\ref{sec:general} are not particularly suitable for this task. The reason is that $\hat{f}_0$ yields an expansion of $y$ around imaginary-order Bessel $K$-function background (recall that the Laplace transform of $\hat{f}_0$ is a divergent and infinitely fast oscillating function at the origin proportional to $K_\nu(x)/\sqrt{x}$ with $\nu=i\sqrt{3}/2$, cf. Eq.~(\ref{eq:KBorelexpression})). Hence, the regular pattern of the local expansion around $x=0$ with $y\to 1$ does not emerge perturbatively in such a scheme, see also~\cite{Dunne:2026hfx}.
%================================
\subsection{Dressing the $\hat{f}_0$ propagator}
Remarkably enough, there is a trick (corresponding, qualitatively, to a dressing of the  hypergeometric ``propagator'' $\hat{f}_0$) that makes it possible to reformulate the problem as a perturbative expansion around another, non-trivial, non-perturbative analytic (and, as we shall see, mathematically very beautiful) background, that may even be viewed as a universal ``template'' of the profiles of `t Hooft-Polyakov monopoles with arbitrary $\beta$, cf.~\cite{MalinskyFuture}. The key is a simple deformation of the $L_\infty[f]$ operator of Eq.~(\ref{eq:foperator}),
\begin{equation}
L_\infty\to L_\infty-\tfrac{3}{x^2}\equiv L^R_\infty\,,
\end{equation}
which (upon $y\to x e^{-x} f$) makes it possible to transform the original ODEs for $f$, i.e., $L_\infty[f](x)=e^{-2x}f(x)^3$ in the MNBPS case equivalent to Eq.~(\ref{eq:MNBPSprofile}), or Eq.~(\ref{eq:VolterraGeneral}) in the NBPS setting equivalent to Eq.~(\ref{eq:ODEGeneralfg}) with only the leading RHS contribution retained, to the form
\be\label{eq:ODEtildef}
L^R_\infty[f^R]=\frac{e^x}{x}+e^{-2x}{f^R}(x)^3-3\frac{e^{-x}}{x}{f^R}(x)^2\,,
\ee 
where $f^R\equiv f-\tfrac{e^x}{x}$. 
The power of this rearrangement stems from the fact that the Volterra equation equivalent to Eq.~(\ref{eq:ODEtildef}) thus assumes a rather special form  
\begin{equation}\label{eq:superVolterra}
t(t+2)\hat{f}^R(t)-2(t+1)\int_0^t {\rm d}s\,\hat{f}^R(s)=\hat{R}(t)\,,
\end{equation}
where $\hat{R}$ is the Borel-plane equivalent of the RHS of Eq.~(\ref{eq:ODEtildef}), 
\be\label{eq:hatR}
\hat{R}(t)=1_{-1}(t)+(\hat{f}^R*\hat{f}^R*\hat{f}^R)(t-2)-3(\hat{f}^R*\hat{f}^R*1)(t-1)\,.
\ee 
Note in particular that the homogeneous solution of Eq.~(\ref{eq:superVolterra}) defining the fundamental mode of this expansion is very simple, namely
\be
\hat{h}(t)=C(t+1)\,,
\ee          
and the solution to the full Eq.~(\ref{eq:superVolterra}) can be thus formally written as
\be\label{eq:hattildef}
\hat{f}^R(t)=\frac{\hat{R}(t)}{t(t+2)}+2(t+1)\int_0^t{\rm d}s\frac{\hat{R}(s)}{s^2(s+2)^2}\,.
\ee
The expansion $\hat{f}^R=\hat{f}^R_0+\hat{f}^R_1+\ldots$ with each $\hat{f}^R_n$ given by Eq.~(\ref{eq:hattildef}) with a suitable part $\hat{R}_n$ of the total $\hat{R}=\hat{R}_0+\hat{R}_1+\ldots$, can then be viewed as and expansion around a non-perturbative background profile $\hat{f}^R_0$ corresponding to $\hat{R}_0=1_{-1}$ (unity anchored at $t=-1$) with the perturbative modes $\hat{f}^R_n$ ($n\in \mathbb{N}$) defined by the triangular partition of the non-linear part of $\hat{R}$ of formula~(\ref{eq:hatR}) corresponding to the expansion of $\hat{f}^R$.
The key point is that the solution~(\ref{eq:hattildef}) of Eq.~(\ref{eq:superVolterra}) with $\hat{R}_0$ on the RHS is very simple (in $u=t+1$, the local coordinate around $t=-1$), namely,
\be
\hat{f}^R_0(u)_{-1}=\frac{1+u_0(u\!-\!u_0)}{u_0^2-1}+u\left[{\rm arctanh}(u)\!-\!{\rm arctanh}(u_0)\right],
\ee 
where $u_0$ is an integration constant to be fixed by one of the boundary conditions. Interestingly, $u_0=0$ yields
\be\label{eq:resultingf0}
\hat{f}^R_0(u)_{-1}= u\;{\rm arctanh}(u) -1
\,,
\ee 
which, upon Laplace transform (localized at $u$), gives
\be
f^R_0(x)=\frac{1}{2 x^2}\left[e^{2x}(x-1){\rm Ei}(-x)+(x+1){\rm Ei}(x)\right]\,.
\ee 
Crucially, $y_0$ derived from this structure obeys both boundary conditions $y_0|_{x\to 0}\to 1$ and $y_0|_{x\to \infty} \to 0$ {\em simultaneously}\,! 
At first glance, this looks like a pure serendipity, but it is actually the intended consequence of the very specific reparametrization of $\hat{f}$ in terms of the resummed quantities. 
On a similar footing (cf.~\cite{MalinskyFuture}), one obtains $g^R_0(x)=[1-\exp(-\sqrt{2\beta}x)(1+\sqrt{2\beta}x)]/\beta x^2$, that corresponds to the Borel-plane structure $\hat{g}^R_0(t)=-t/\beta+\theta(t-\sqrt{2\beta})\left[1+t-\sqrt{2\beta}\right]$ localized at $0$ (global algebraic germ) and $\sqrt{2\beta}$ (shifted exponential sector germ), respectively.
%=========================================================
\subsection{Structure of the resummed expansion}
\begin{figure}[t]
  \includegraphics[width=0.43\textwidth]{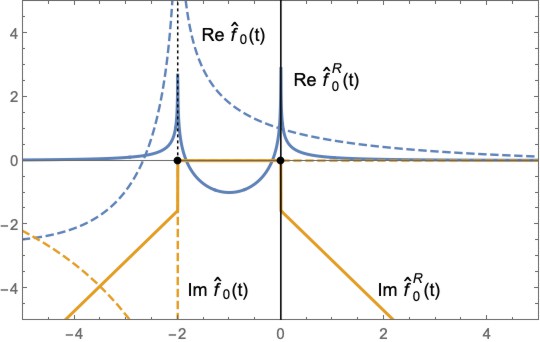}\;
\caption{\label{fig:propagators}
Borel plane structure of the ``naked'' propagator $\hat{f}_0$ (dashed) of Secs.~\ref{sec:MNBPS} and~\ref{sec:general} and its dressed counterpart $\hat{f}^R_0$ of Sec.~\ref{sec:resummation} (solid); real parts in blue, imaginary in yellow. One can see the emergence of the second branch cut of $\hat{f}_0^R$ at $t=0$ due to the partial resummation of $\hat{f}$. Note also the fast diminishing of the asymptotic tails of $\hat{f}_0^R$, as compared to $\hat{f}_0$.} 
\end{figure}

Interestingly, switching from the na\"\i ve hypergeometric expansion seed $\hat{f}_0(t)\propto{}_2F_1(\tfrac{1}{2}-i\tfrac{\sqrt{3}}{2},\tfrac{1}{2}+i\tfrac{\sqrt{3}}{2};1;-\tfrac{t}{2})$ to the nontrivial background $\hat{f}^R_0$ of Eq.~(\ref{eq:resultingf0}), the Borel-plane structure of the perturbative expansion of $\hat{f}^R$ (and $\hat{g}^R$) also changes with respect to that of $\hat{f}$ and $\hat{g}$ discussed in Sects.~\ref{sec:MNBPS} and~\ref{sec:general}. Note in particular that the ``resummed'' vector seed profile $\hat{f}^R_0$, which shares the $t=-2$ branch-cut of its ``naked'' variant $\hat{f}_0$, develops a new singularity at $t=0$ where $\hat{f}_0$ is regular, cf. Fig.~\ref{fig:propagators}. Second, both real and imaginary parts of $\hat{f}^R_0$ are very nicely behaved in the large-$t$ region where the real part of $\hat{f}_0$ suffered from oscillatory behaviour with roots at exponentially distant points. Third, unlike in $\hat{f}_0$, no free constant remains in $\hat{f}^R_0$. This is a clear indication of a significant qualitative difference of the two types of perturbative expansions and, in fact, of the superiority of the latter. 

Moreover, since $\hat{f}^R_0$ on its own generates an $x$-plane profile $y_0$ that obeys both boundary conditions, higher-order corrections do not need to produce any non-perturbative jump at $x=0$ like it was required from the expansion around $\hat{f}_0$. Hence, they may not only vanish at the $x=0$ singularity and naturally decay at $x\to \infty$, but they can gradually diminish on the entire $\mathbb{R}_+$; thus, the resummed expansion may even converge uniformly to the exact solution. This is demonstrated explicitly in~\cite{MalinskyFuture}.          
   
%=========================================================
\subsection{Sample lowest-order $B_\infty$ calculation\label{sec:Bparam}}
Finally, $y_0$ as the non-perturbative background of the full perturbative expansion of $y$ behaves around $x=0$ as
\be
y_0(x)=1+\tfrac{1}{3}x^2(\gamma_E-\tfrac{4}{3}+\log x)+\ldots\,,
\ee
where $\gamma_E$ is the Euler-Mascheroni constant.
Thus, one verifies the structure of the leading $x^2 \log x$ term of the expansion~(\ref{eq:yexpansionaround0}), and obtains the lowest-order part of $B_\infty$,
\be
B_{\infty,0}=\tfrac{1}{3}(\gamma_E-\tfrac{4}{3})\sim -0.252\,.
\ee
This is in the right ballpark of the ``true'' value $B_\infty\sim -0.484$ obtained from the numerics, cf. Appendix~\ref{app:MNBPStransseries}. Higher order corrections, along with an {\em exact analytic formula for $B_\infty$}, are given in the extended study~\cite{MalinskyFuture}.

\section{Conclusions}\label{sec:conclusions}
In this letter we have elucidated the remarkable Borel-plane structure of the solutions of the 't~Hooft-Polyakov monopole profile equations for all positive values of the $\beta=\lambda/e^2$ parameter therein. It was shown that, within the canonical perturbative scheme suggested by the universal asymptotics of the monopole gauge profiles, the emerging discrete singularity pattern is fully driven by the ${}_2F_1(\tfrac{1}{2}-i\tfrac{\sqrt{3}}{2},\tfrac{1}{2}+i\tfrac{\sqrt{3}}{2};1;-\tfrac{t}{2})$ hypergeometric function featuring a logarithmic branch cut at $t=-2$. We argued that a substantially better universal global expansion scheme can be devised around a non-trivial non-perturbative background featuring a surprisingly simple analytic structure (explicitly constructed for the vector sector), later used as a basis for a sample calculation of the numerical parameter $B_\infty$ governing the local $x=0$ expansion of the gauge-sector profile in the MNBPS $\beta\to\infty$ setting.

Hence, the Borel-plane perspective looks like a very natural viewpoint for contemplating topological defects in spontaneously broken gauge theories.

\section*{Acknowledgments}
The work has been performed with the support from the 
Charles University Research Center of Excellence 
UNCE/24/SCI/016 grant and from the FORTE project CZ.02.01.01/00/22\_008/0004632 co-funded by the EU and the Ministry of Education, Youth and Sports of the Czech Republic. The author is grateful to Petr Bene\v{s} and Filip Blaschke for valuable feedback on the manuscript. The paper is dedicated to the memory of the author’s mother.
\appendix
%-------------------------------------------
\section{Universal $\beta>0$ gauge asymptotics}\label{app:y0MNBPS}
The asymptotic profiles of the solutions of Eq.~(\ref{eq:gaugeprofile}) correspond to a mere replacement of $z$ by its asymptotic limit $z(x)=1$ which, for $\beta>0$, is saturated exponentially fast, i.e. $1-z(x) \sim e^{-2x}$ for large $x$ (unlike in the BPS case where the convergence is only hyperbolic, $1-z(x) \sim x^{-1}$). With the fast-decaying $y^3/x^2$ piece neglected, the profile equation~(\ref{eq:gaugeprofile}) assumes a simple linearized form
\begin{equation}\label{eq:linearizedMNBPS}
y''(x)=\left(1-\tfrac{1}{x^2}\right)y(x)\,.
\end{equation}
For $y(x)\equiv\sqrt{x}\,u(x)$ this structure can be readily mapped onto the defining equation of Bessel functions 
$
x^2 u''(x)+x u'(x)-(x^2+\nu^2)u(x)=0
$,
with $\nu^2=-3/4$ and, as such, the physically interesting (i.e. decaying) universal asymptotic solution of Eq.~(\ref{eq:gaugeprofile}) can be written as
$
A \sqrt{x}K_\nu(x)
$, 
where $K_\nu$ is the modified Bessel $K$-function of imaginary order $\nu=i\sqrt{3}/2$ and $A$ is a real constant. 

%-------------------------------------------
\section{The MNBPS $y(x)$ transseries at $x\to\infty$}\label{app:MNBPStransseriesInfinity}
The large-$x$ transseries for $y$ in the $\beta\to\infty$ limit obtained from an ansatz
\begin{equation}\label{eq:InfinityTransseries}
y(x)= \sum_{m=0}^\infty \sum_{n=0}^m  a_{m,n}x^{-m}{e^{-(2n+1)x}}
\end{equation}
by a mere order-by-order comparison of the coefficients at the LHS and RHS of Eq.~(\ref{eq:MNBPSprofile}) clearly reveals its asymptotic nature in the numerical behaviour of $a_{m,n}$; for instance, the $n=0$ tower (stripped from an a-priori unknown overall factor $A_\infty$) reads
\begin{displaymath}\label{eq:acoefficients}
a_{m,0}\propto\{1,\;-\tfrac{1}{2},\;\tfrac{3}{8},\;-\tfrac{7}{16},\;\tfrac{91}{128},\;-\tfrac{1911}{1280},\;\tfrac{19747}{5120},\;
\ldots\}\,.
\end{displaymath}
As expected, these are (up to the standard Borel-transform factorials) exactly the coefficients of the power expansion of ${}_2F_1(\tfrac{1}{2}-i\tfrac{\sqrt{3}}{2},\tfrac{1}{2}+i\tfrac{\sqrt{3}}{2};1;-\tfrac{t}{2})$ around $t=0$.   
%------------------------------------------
\section{The MNBPS $y(x)$ transseries at $x=0$\label{app:MNBPStransseries}}
Similarly, the behaviour of the gauge profile of the MNBPS ($\beta\to\infty$) monopole around $x=0$ can be written in terms of a locally convergent power-log transseries 
\begin{equation}\label{eq:yexpansionaround0}
y(x)=\sum_{m=0}^\infty \sum_{n=0}^m b_{m,n}x^{2m}\log^n x\,,
\end{equation}
whose $b_{m,n}$ coefficients can be fully determined recursively from the $y|_{x\to 0}\to 1$ initial condition, up to one a-priori unknown constant $B_\infty$. The first few terms of this expansion read (see also~\cite{Forgacs:2005vx})
\begin{align}\label{eq:bcoefficients}
&y(x)=1+x^2(B_\infty+\tfrac{1}{3}\log x)+x^4\left[\tfrac{1}{30}(\log x)^2+\right.\\
&+\left.(\tfrac{1}{5}B_\infty-\tfrac{1}{75})\log x+\tfrac{1}{750}(225 B_\infty^2-30 B_\infty+2)\right]+\ldots\nonumber\,,
\end{align}
with a numerical fit (see~\cite{MalinskyFuture}) revealing $B_\infty\sim -0.484$. This series has a finite convergence radius $R$ that can be bounded from above, for instance, by the progression of its leading-log coefficients $b_{n,n}$ that, asymptotically, is purely geometrical; for large $n$, $b_{n,n}\approx \sqrt{8}\,q^n$ with $q\approx~0.109374$. Hence, $R\lesssim \exp[\tfrac{1}{2}W\left({2}/{q}\right)]\sim 2.92$, where $W$ is the Lambert function.
%\bibliographystyle{h-physrev5}
%\bibliography{bibliography-mnbps}

\end{document}